\documentclass{article}

\usepackage{mathrsfs}
\usepackage{arxiv}
\usepackage{amsfonts}
\usepackage[utf8]{inputenc} 
\usepackage[T1]{fontenc}    
\usepackage{hyperref}       
\usepackage{url}            
\usepackage{booktabs}       
\usepackage{amsfonts}       
\usepackage{nicefrac}       
\usepackage{microtype}      
\usepackage{lipsum}
\usepackage{graphicx}
\usepackage{amsmath}
\graphicspath{ {./images/} }

\title{Review for Dynamic prediction in Clinical Survival Analysis}

\author{
 Weiyi He \\
  University of Science and Technology of China\\
  Hefei, China\\
  \texttt{hwyii@mail.ustc.edu.cn} \\
}

\begin{document}
\maketitle
\begin{abstract}
The accurate prediction of patient prognosis is a critical challenge in clinical practice. With the availability of various patient information, physicians can optimize medical care by closely monitoring disease progression and therapy responses. To enable better individualized treatment, dynamic prediction models are required to continuously update survival probability predictions as new information becomes available. This article aims to offer a comprehensive survey of current methods in dynamic survival analysis, encompassing both classical statistical approaches and deep learning techniques. Additionally, it will also discuss the limitations of existing methods and the prospects for future advancements in this field.
\end{abstract}


\section{Introduction}
Dynamic prediction in survival analysis means (i). Utilizing the entire longitudinal history data (ii). Updating the predictions each time new data becomes available. It employs the whole history of the longitudinal measurements instead of just a small period of information in the classical survival analysis model. This approach is vital in various fields, including healthcare and finance, where the ability to adapt to changing circumstances is crucial for accurate predictions. By continuously incorporating new information, dynamic prediction enables better decision-making and improves the accuracy of prognostic assessments.

Previous research in this domain has explored various methods related to dynamic prediction. For example, \cite{van2011dynamic} introduce the dynamic prediction by focusing on different datasets. However, existing reviews have mostly focused on statistical approaches, overlooking the deep learning methods.

As the field of deep learning continues to evolve, there is potential for its application in dynamic prediction. Therefore, in this review, we aim to bridge the gap between statistical and deep learning methods for solving dynamic prediction problems in survival analysis. We present a comprehensive overview of approaches from both statistical and deep learning domains, exploring their strengths and limitations for dynamic prediction tasks. Moreover, we directly discuss the difference among various covariates and longitudinal outcomes.

\section{Background}
\label{sec:headings}
\subsection{Survival analysis}
Let $T^*$ denote the random variable of failure times under study. 
Let $D_n=\{T_i,\delta_i,\boldsymbol{y}_i;i=1,\cdots,n \}$ denote a sample from the population, where $T_i=\operatorname{min}(T_i^*,C_i)$ denotes the observed event time for the $i$-th subject $(i=1,\cdots,n)$ with $T_i^*$ denoting the true event time, $C_i$ the censoring time, and $\delta_i=I(T_i^*\leq C_i)$ the event indicator.\\
Survival function is used to describe the distribution of $T^*$. It expresses the probability that the event occurs after $t$, that is the probability of surviving time $t$. It is defined as
\begin{equation}
S(t)=\Pr(T^*>t)=\int_t^{\infty}p(s)ds, 
\end{equation}
where $p(\cdot)$ denotes the corresponding probability density function. \\
The hazard function describes the instantaneous risk for an event in the time interval $[t,t+dt)$ provided survival up to $t$, and is defined as
\begin{equation}
h(t)=\lim_{t \to 0} \frac{\Pr(t\leq T^* < t+dt|T^*\ge t)}{dt},t>0.
\end{equation}
The survival also can be expressed in terms of the risk function as
\begin{equation}
S(t)=\operatorname{exp}\{-H(t)\}=\operatorname{exp}\{-\int_0^th(s)ds\},
\end{equation}
where $H(\cdot)$ is known as the cumulative risk function that describes the accumulated risk up until time $t$.
\subsection{Dynamic prediction}
In dynamic prediction setting, we are interested in utilizing all available information at hand (including both baseline information and accumulated biomarker levels) to produce predictions of survival probabilities. To put it formally, we are interested in predicting survival probabilities for a new subject $i$ that has provided a set of longitudinal measurements $Y_i(t) = \{y_i(s); 0 \leq s < t\}$. The fact that biomarker measurements have been recorded up to $t$, implies survival of this subject up to this time point, meaning that it is more relevant to focus on the conditional subject-specific predictions, given survival up to $t$. Hence, it is more relevant to focus on the conditional probability of surviving time $u > t$ given survival up to $t$, i.e.,
\begin{equation}
\pi_i(u|t)=\Pr(T_i^*\ge u|T_i^*>t,Y_i(t),D_n).
\end{equation}
The dynamic prediction for $\pi_i(u|t)$ is evident since when new information is recorded for patient $i$ at $t'>t$, we can update the prediction to obtain $\pi_i(u|t')$ with $u>t'$, therefore to proceed in a time-dynamic manner. 
\subsection{Covariates and Longitudinal Outcomes} 
We introduce the definition of three common covariates and longitudinal outcomes in the setting of survival analysis.
\paragraph{Baseline Covariates}

Baseline covariates in survival analysis are fixed and do not change over time for each individual.
It is usually defined as a qualitative factor or a quantitative variable measured or observed before a subject starts taking study medication (usually before randomisation) and expected to influence the primary variable to be analysed.

\paragraph{External(Exogenous) Covariates}
The most important characteristic of external covariates is that the value of these covariates at any time point $t$ is not affected by the true failure time. That is, no matter the failure happen or not, the future path of the covariate is not affected. In mathematical form, they should satisfy
\begin{equation}
\Pr\{Y_i(t)|Y_i(s),T_i^*\ge s\}=\Pr\{Y_i(t)|Y_i(s),T_i^*=s\},\quad s\leq t.
\end{equation}
The most common examples are stochastic processes that are external to the subjects under study. For instance, the levels of environmental factors, such as air pollution. Other examples are available in \cite{rizopoulos2012joint}

\paragraph{Internal(Endogenous) Covariates}
Internal covariates typically arise as time-dependent measurements taken on the subjects under study. They typically require the survival of the subject for their existence. Thus, when failure is defined as the death of the subject, their path carries direct information about the failure time. That is, only when the subject is alive can we get the internal covariates. Failure of the subject at time $s$ corresponds to nonexistence of the covariate at $t \ge s$, which has as a direct implication the violation of the condition (5).

In summary, the main purpose of including covariates in survival analysis is to investigate how these variables impact the hazard of the event (e.g., death) while accounting for their influence on the outcome.

\paragraph{Longitudinal Outcomes}

Longitudinal outcomes refer to repeated measurements of a variable over time for each individual in a study. These repeated measurements are often collected at specific time points during the follow-up period. Longitudinal outcomes capture the changes in the variable of interest over time and can provide insights into the trajectory or progression of that variable.

To our best knowledge, there is some overlaps in the specific referents of longitudinal outcomes and internal covariates; they provide explanations from two different perspectives.

\section{Methods}
Here we summarize state-of-art approaches for dynamic prediction in clinical survival analysis. Previous literature [e.g. \cite{van2011dynamic}, \cite{rizopoulos2017dynamic}] only included the basic statistics models such as landmark methods and joint models. Instead, we provide a more comprehensive review which covers the extensions of those statistic models as well as some recent contributions made by the deep learning community.
\subsection{Joint modeling}
The extended Cox model is only appropriate for exogenous time-dependent covariates and therefore cannot handle longitudinal biomarkers \cite{rizopoulos2012joint}
. When primary interest is in the association between such endogenous time-dependent covariates and survival, an alternative modeling framework has been introduced in the literature, known as the joint modeling framework for longitudinal and time-to-event data. Under this framework, for a specific patient and at a specific time point during follow-up, we can easily utilize all available information we have at hand to produce predictions of survival probabilities.
\subsubsection{Basic Joint models}
\paragraph{Submodels}
Using a linear mixed-effects model to describe the subject-specific longitudinal trajectories. Namely, for subject $i$, we have
\begin{equation}
\begin{aligned}
y_i(t)&=m_i(t)+\epsilon_i(t)=\boldsymbol{x}_i^T(t)\boldsymbol{\beta}+\boldsymbol{z}_i^T(t)\boldsymbol{b}_i+\epsilon_i(t),\\
\boldsymbol{b}_i &\sim \mathscr{N}(\boldsymbol{0},\boldsymbol{D}),\epsilon_i(t) \sim \mathscr{N}(0,\sigma^2), 
\end{aligned}
\end{equation}
where $y_i(t)$ denotes the value of the longitudinal outcome at any particular time point $t$, $\boldsymbol{x}_i(t)$ and $\boldsymbol{z}_i(t)$ denote the time-dependent design vectors for the fixed-effects $\boldsymbol{\beta}$ and for the random effects $\boldsymbol{b}_i$, respectively, and $\epsilon_i(t)$ the corresponding error terms that are assumed independent of the random effects, and $\operatorname{cov}\{\epsilon_i(t),\epsilon_i(t')\}=0$ for $t'\neq t$. For the survival process, we assume that the risk for an event depends on the true value of the marker at time $t$, denoted by $m_i(t)$. More specifically, we have
\begin{equation}
h_i(t|M_i(t),\boldsymbol{w}_i)=h_0(t)\operatorname{exp}\{\boldsymbol{\gamma}^T\boldsymbol{w}_i+\alpha m_i(t)\},t>0
\end{equation}
where $M_i(t)=\{m_i(s),0\leq s < t\}$ denotes the history of the true unobserved longitudinal process up to $t$, $h_0(\cdot)$ denotes the baseline hazard function, and $\boldsymbol{w}_i$ is a vector of baseline covariates with corresponding regression coefficients $\boldsymbol{\gamma}$. To make subject-specific predictions, we need to make appropriate assumptions for the baseline hazard function $h_0(\cdot)$ \cite{rizopoulos2012joint}.

Moreover, it is also necessary to explain about random effects and the fix-effects. They are important concepts in statistics, and it turns out that different definitions are used in different contexts\cite{gelman2005analysis}. In our context, fixed effects are constant across individuals, and random effects vary. To be specific, we define effects
(or coefficients) in a multilevel model as constant if they are identical for all groups in a population and varying if they are allowed to differ from group to group. For example, the model $y_{ij} = \alpha_j +\beta x_{ij}$ (of units $i$ in groups $j$) has a constant slope and varying intercepts, and $y_{ij} = \alpha_j + \beta_j x_{ij}$ has varying slopes and intercepts.

\paragraph{Estimation}
The main estimation method that has been proposed for joint models is (semiparametric) maximum likelihood \cite{wulfsohn1997joint}, \cite{henderson2000joint}. Bayesian estimation of joint models using MCMC techniques has been considered by \cite{hanson2011predictive}, \cite{chi2006joint}, \cite{xu2001joint}, \cite{wang2001jointly}, among others. Moreover, \cite{tsiatis2001semiparametric} have proposed a conditional score approach in which the random effects are treated as nuisance parameters, and they developed a set of unbiased estimating equations that yields consistent and asymptotically normal estimators. We give the maximum likelihood method for the joint models as one of the most traditional approaches. 

The likelihood of the models is derived under the assumption that the vector of time-independent random effects $\boldsymbol{b_i}$ accounts for all interdependencies between the observed outcomes. That is, given the random effects, the longitudinal and event time process are assumed independent, and in addition, the longitudinal responses of each subject are assumed independent.
Formally, we have
\begin{equation}
\begin{aligned}
p(T_i,\delta_i,y_i|\boldsymbol{b_i};\theta)&=p(T_i,\delta_i|\boldsymbol{b_i};\theta)p(y_i|\boldsymbol{b_i};\theta),\\
p(y_i|\boldsymbol{b_i};\theta)&=\prod_{j}p\{y_i(t_{ij})|\boldsymbol{b_i};\theta\}
\end{aligned}
\end{equation}
Then we can train the model by maximizing log-likelihood function for the $i$th subject.
\begin{equation}
\begin{aligned}	\operatorname{log}p(T_i,\delta_i,y_i;\theta)&=
	\operatorname{log}\int p(T_i,\delta_i,y_i,\boldsymbol{b_i};\theta)d\boldsymbol{b_i}\\
	&=\operatorname{log}\int p(T_i,\delta_i|\boldsymbol{b_i};\theta_t,\beta)[\prod_{j}p\{y_i(t_{ij})|\boldsymbol{b_i};\theta_y\}]p(\boldsymbol{b_i};\theta_b)d\boldsymbol{b_i}
\end{aligned} 
\end{equation}
where
\begin{equation}
\begin{aligned}
p(T_i,\delta_i|\boldsymbol{b_i};\theta_t,\beta) &= h_i(T_i|M_i(T_i);\theta_t,\beta)^{\delta_i}S_i(T_i|M_i;\theta_t,\beta)\\
	&= [h_0(T_i)\operatorname{exp}\{\gamma^Tw_i+\alpha m_i(T_i)\}]^{\delta_i}\\
	&\times \operatorname{exp}(-\int_{0}^{T_i}h_0(s)\operatorname{exp}\{\gamma^Tw_i+\alpha m_i(s)\}ds)
\end{aligned}
\end{equation}

\begin{equation}
\begin{aligned}
p(y_i|\boldsymbol{b_i};\theta)p(\boldsymbol{b_i};\theta)&=\prod_{j}p\{y_i(t_{ij})|\boldsymbol{b_i};\theta_y\}p(\boldsymbol{b_i};\theta_b)\\
&= (2\pi \sigma^2)^{-n_i/2}\operatorname{exp}\{-||y_i-X_i\beta-Z_i\boldsymbol{b_i}||^2/2\sigma^2\}\\
&\times (2\pi)^{-q_b/2}\operatorname{det}(D)^{-1/2}\operatorname{exp}(-\boldsymbol{b_i}^TD^{-1}\boldsymbol{b_i}/2)
\end{aligned}
\end{equation}

\paragraph{Prediction}
Under the assumption in the last section, we have
\begin{equation}
\begin{aligned}
Pr(T_i^*\ge u|T_i^*>t,Y_i(t);\theta)&=\int_{}^{} Pr(T_i^*\ge u|T_i^*>t,\boldsymbol{b_i};\theta)p(\boldsymbol{b_i}|T_i^*>t,Y_i(t);\theta)d\boldsymbol{b_i} \\
	&= \int_{}^{}\frac{S_i\{u|M_i(u,\boldsymbol{b_i},\theta);\theta\}}{S_i\{t|M_i(t,\boldsymbol{b_i},\theta);\theta\}}p(\boldsymbol{b_i}|T_i^*>t,Y_i(t);\theta)d\boldsymbol{b_i}
\end{aligned}  
\end{equation}
where
\begin{equation}
\begin{aligned}
 S_i(t|M_i(t),\boldsymbol{w_i})&= 
 Pr(T_i^*>t|M_i(t),\boldsymbol{w_i})\\
 	&=\operatorname{exp}\left( -\int_{0}^{t}h_0(s)\operatorname{exp}\{\gamma^T\boldsymbol{w_i}+\alpha m_i(s)\}ds\right)
\end{aligned}
\end{equation}
denotes the subject-specific survival function.

The prediction formula in (4) can be written as
\begin{equation}
\begin{aligned}
\pi_i^{JM}(u|t)&=\int \Pr(T_i^*\ge u,\boldsymbol{\theta}|T_i^*>t,Y_i(t),D_n)d\boldsymbol{\theta}\\
&=\int Pr(T_i^*\ge u|T_i^*>t,Y_i(t),\boldsymbol{\theta})p(\boldsymbol{\theta}|D_n)d\boldsymbol{\theta}
\end{aligned}
\end{equation}
The first part of the integrand is given by (11). For the second part, which is the posterior distribution of the parameters given the observed data, we assume that the sample size $n$ is sufficiently large such that $\{\theta|D_n\}$ can be well approximated by $\mathscr{N}\{\hat{\theta},\hat{\operatorname{var}}(\hat{\theta})\}$.
Then we can derive a Monte Carlo estimate of $\pi_i(u|t)$.\cite{rizopoulos2012joint} The implementation of the prediction is given in \cite{rizopoulos2010jm}.

In the next section, we will give some common extensions of the basic joint model. To better handle real-world datasets, we can consider exogenous covariates, competing risks, and multiple longitudinal outcomes. One can get comprehensive extensions and their examples in \cite{rizopoulos2012joint}, with the application in $R$. In fact, R packages \textbf{JM} provide function \textit{jointModel()} to help fit the models.

\subsubsection{Exogenous Time-Dependent Covariates}
It is natural to consider both the internal and external time-dependent covariates in the model. For example, the levels of environmental factors, such as air pollution, may be associated with the frequency of asthma attacks. The relative risk model in (7) can be easily extended to handle exogenous covariates, as
\begin{equation}  h_i(t)=h_0(t)\operatorname{exp}\{\gamma^Tw_i(t)+\alpha m_i(t)\},
\end{equation}
where the covariate vector $w_i(t)$ now contains both baseline and exogenous time-dependent covariates.\cite{rizopoulos2012joint} The estimation of the this model proceeds in the same manner as the basic joint model. The only difference is for the definition of the survival function, which is expanded as
\begin{equation}
\begin{aligned}
S_i(t|M_i(t),w_i)&=\operatorname{exp}\left (-\int_{0}^{t}h_i(s)ds \right )\\&=
\operatorname{exp} \left ( 
-\sum_{q=1}^{Q_i}\int_{\Omega_{iq}}h_0(s)\operatorname{exp}\{\gamma^Tw_{iq}(s)+\alpha m_i(s)\}ds
\right ),
\end{aligned}
\end{equation}
where $\{\Omega_{iq},q=1,\cdots,Q_i\}$ denote the time intervals during which the exogenous time-dependent covariates $w_i(t)$ are assumed constant.

\subsubsection{Multiple Longitudinal Outcomes} 
The basic joint model handle only one longitudinal outcome. However, it is common to deal with multiple longitudinal outcomes. The extension of joint models to handle more than one longitudianl outcome is straightforward under the Generalized Linear Mixed Model(GLMMs) framework \cite{rizopoulos2012joint}. \cite{song2002estimator}, \cite{chi2006joint} also extended the longitudinal model to the multivariate case.

\subsubsection{Competing risks} 
In some cases, a subject can experience more than one type of mutually exclusive events — typically referred to as competing risks (CR). For instance, a patient can die from different causes (e.g. cancer or non-cancer death). If the main focus is a specific event type, others could be recorded as censored observations. Joint models of competing risks have been studied by \cite{williamson2008joint}, \cite{elashoff2008joint}.
In particular, assuming $K$ different causes of failure, let $T_{i1}^*,\cdots,T_{iK}^*$ denote the true failure times for each. Then the observed event time can be renewed as $T_i=\operatorname{min}(T_{i1}^*,\cdots,T_{iK}^*,C_i)$, and the event indicator takes values $\delta_i\in \{0,1,\cdots,K\}$, with $0$ corresponding to censoring, and $1,\cdots,K$ to the competing events. Therefore, for each of the $K$ CR, 
\begin{equation}
    h_{ik}(t)=h_{0k}(t)\operatorname{exp}\{\gamma_k^Tw_i+\alpha_k m_i(t)\}
\end{equation}
For the estimation of this kind of joint model, we only need to modify (10) as
\begin{equation}
\begin{aligned}
p(T_i,\delta_i|\boldsymbol{b_i};\theta_t,\beta)
&= \prod_{k=1}^{K}[h_{0k}(T_i)\operatorname{exp}\{\gamma_k^Tw_i+\alpha_k m_i(T_i)\}]^{I(\delta_i=k)}\\
	&\times \operatorname{exp}\left(-\sum_{k=1}^{K}\int_{0}^{T_i}h_{0k}(s)\operatorname{exp}\{\gamma_k^Tw_i+\alpha_k m_i(s)\}ds\right) 
\end{aligned}
\end{equation}

\subsubsection{BMA}
Bayesian model averaging (BMA)\cite{hoeting1999bayesian} can be used for joint models to do the dynamic prediciton. The novel feature of this approach is that the weights for combing predictions depend on the recorded information for the subject for whom predictions are of interest. Thus, for different subjects and even for the same subject but at different follow-up times, different models may have higher weights. This explicitly accounts for model uncertainty and acknowledges that a single prognostic model may not be adequate for quantifying the risk of all patients. \cite{rizopoulos2014combining}
Let $D_j(t)=\{Y_j(t),T_j^*>t,\boldsymbol{w}_j\}$ denote the available data for this subject. The survival probability is
\begin{equation}
\Pr(T_j^*>u|D_j(t),D_n)=\sum_{k=1}^{K}\Pr(T_j^*>u|M_k,D_j(t),D_n)p(M_k|D_j(t),D_n).
\end{equation}
where $M_1,\cdots, M_K$ joint models with different association structures.

\subsection{Landmark methods}
The basic idea behind landmarking is to obtain survival probabilities from a Cox model fitted to the patients from the original dataset who are still at risk at the time point of interest (e.g., the last time point we know that the new patient was still alive). As in \cite{van2007dynamic}, we focus on using landmarking approach to obtain predictive probabilities for the case of time-dependent covariates.
\subsubsection{Standard Landmarking}
In the case of a time-dependent covariate the clinical interest is in the prediction of survival given the covariate history up to a dynamic landmark point. The approach is through direct modelling using the landmark paradigm.
The simplest landmark model is \cite{van2007dynamic}
\begin{equation}
h(t|X(s))=h_0(t|s)\operatorname{exp}(X(s)\beta_{LM}(s)),t\ge s.
\end{equation}
The point is to obtain a simple model for $h_0(t|s)$ and $\beta_{LM}(s)$.
They present the pseudo-likelihoods to fit models for $\beta_{LM}(s)$. The way is inspired by fitting the models in standard software by creating new enlarged data sets.

The estimate of the (landmark specific) baseline hazard is given by 
\begin{equation}
\hat{h}_0(t_i|s)=\frac{1}{\sum_{t_j\geq t_i}\operatorname{exp}(X_j(s)\hat{\beta}_{LM}(s)}),t_i>s
\end{equation}
\subsubsection{Competing risks}
Similar to the extension of the joint model, \cite{nicolaie2013dynamic} propose an extension of the landmark model to the problem of dynamic prediction in competing risks with time-dependent covariates. To be specific, they fix a set of landmark time points $t_{LM}$ within the follow-up interval. For each of these landmark time points $t_{LM}$, they create a landmark data set by selecting individuals at risk at $t_{LM}$ and fix the value of the time-dependent covariate in each landmark data set at $t_{LM}$. Then they assume Cox proportional hzard models for the cause-specific hazards and consider smoothing the (possibly) time-dependent effect of the covariate for the different landmark data sets.

\subsection{Deep learning methods}
Conventional statistics methods are often constrained by strong parametric assumptions and limited in their ability to learn from high-dimensional data. With the development of deep learning method, we can find that deep networks always achieved significantly improved performance in survival analysis owing to the ability to represent complicated associations between features and outcomes.

\subsubsection{Dynamic-DeepHit}
Changhee has proposed Dynamic-DeepHit \cite{lee2019dynamic} that extends their previous work in \cite{Lee_Zame_Yoon_Schaar_2018} to dynamic survival analysis, incorporating longitudinal measurements of biomarkers and risk factors into a model rather than discarding valuable information recorded over time. It can also be used for competition risks. Dynamic-DeepHit is a multi-task network, which consists of two types of subnetworks: a shared subnetwork that handles the history of longitudinal measurements and predicts the next measurements of time-varying covariates, and a set of cause-specific subnetworks which estimates the joint distribution of the first hitting time and competing events.

To be specific, the shard subnetwork consists of a RNN structure to handle the longitudinal data and an attention mechanism to unravel the temporal importance of the history of meansurements. Given the history of longitudinal measurements$\boldsymbol{X}$, define the true CIF as
\begin{equation}
F_{k^*}(\tau^*|\boldsymbol{X})=P(T\leq \tau^*,k=k^*|\boldsymbol{X},T>t^*_{J^*}).
\end{equation}
but it is unknown; we utilized the estimated CIF, $\hat{F}_{k^*}(\tau^*|\boldsymbol{X})$ in order to perform dynamic prediction. To be specific, whenever a new measurement is recorded for this subject at time $t>t^*_{J^*}$, we can update (1) in a dynamic fashion. Given a subject with $\boldsymbol{X}$, each output node represents the probability of having event $k$ at time $\tau$, i.e.,$o^*_{k,\tau}=\hat{P}(T=\tau,k=k^*|\boldsymbol{X})$. Therefore, we can define the estimated CIF for cause $k^*$ at time $\tau^*$ as follows:
\begin{equation}
\hat{F}_{k^*}(\tau^*|\boldsymbol{X})=\frac{\sum_{t^*_{J^*}<\tau \leq \tau^*}o^*_{k^*,\tau}}{1-\sum_{k\neq \empty}\sum_{n\leq t^*_{J^*}}o^*_{k,n}}
\end{equation}
which is built upon the condition that this subject has survived up to the last measurement time. Code for Dynamic-DeepHit is available at \href{https://github.com/chl8856/Dynamic-DeepHit}{https://github.com/chl8856/Dynamic-DeepHit}.

\subsubsection{MATCH-Net}
MATCH-Net\cite{jarrett2019dynamic}: a Missingness-Aware Temporal Convolutional Hitting-time Network proposed by Daniel Jarrett, designed to capture temporal dependencies and heterogeneous interactions in covariate trajectories and patterns of missingness. That is, it uses the temporal convolutions in capturing explicit representations of covariate trajectories to make full use of historical information in issuing dynamic predictions and explicitly accounts for informative missingness by learning correlations between patterns of data missingness and diease progression.

Formally, define 
\begin{equation}
\boldsymbol{X}_{i,t,w}=<x_{i,t'}>_{t'\in T},\quad T=\{t':t-w\leq t' \leq t\}
\end{equation}
to be the set of observations for patient $i$ extending from time $t$ into a width-$w$ window of the past.
Given a \textit{backward-looking} historical window $(t-w,t]$, we are interested in the failure function for \textit{forwad-looking} prediction intervals $(t,t+\tau]$; that is, we want to estimate the probability
\begin{equation}
F_i(t+\tau|t,w) = \mathbb{P}(T_{i}\leq t+\tau|T_i>t,\boldsymbol{X}_{i,t,w})
\end{equation}
of event occurrence within each prediction interval. Since the true distribution of survival times cannot be known on the basis of a finite dataset, the objective is therefore to obtain estimates of the true probability.\\
In MATCH-Net architecture, the convolutional block first learns representations of longitudinal covariate trajectories by extracting local features from temporal patterns in the data. Indicator masks are processed in a parallel stream, and filter activations from the auxiliary branch are concatenated with those in the main branch after each layer. The fully-connected block then captures more global relationships by combining local information extracted from the convolutional block. Finally through the network, each prediction task produces the array of failure estimates
\begin{equation}
\hat{\boldsymbol{y}}_{i,t}=[\hat{F}_i(t+1|t,w),\cdots,\hat{F}_i(t+\tau_{max}|t,w)]
\end{equation}
where $\tau_{max}$ is the maximal prediction horizonl; the sequence $\hat{\boldsymbol{y}}_{i,t}$ traces out the survival curve for patient $i$ conditioned on survival until $t$. 

\section{Discussion}

In this study, we have demonstrated that deep learning methods show promising results and can effectively incorporate competing risks into the analysis. However, we acknowledge that there are certain limitations and challenges that need to be addressed. One concern is that deep learning methods tend to combine all kinds of covariates into a single longitudinal history $\boldsymbol{X}$. This may not be appropriate as internal and external covariates possess different properties and require distinct treatment. It is crucial to consider their respective natures and possibly apply different processing techniques to enhance the model's performance.

For the landmark approach, although it has shown some success in dynamic prediction, we must note that it is not fully dynamic as survival predictions are only available at the predefined landmarking times, not at times at which new measurements are obtained. This may raise questions about the generalizability and robustness of the model.

A promising alternative is the Joint model, which allows for distinguishing between different types of covariates and may reduce prediction errors. However, it still maintains a proportional hazard assumption, which might not always hold in real-world scenarios. Furthermore, with the growing availability of electronic medical records, there is an increasing abundance of data, and the prevalence of multi-dimensional longitudinal outcomes is on the rise. In such cases, using traditional joint models could pose significant computational challenges. Moreover, when implementing joint models using R, the forms of random effects and fixed effects in the linear mixed effect part are typically user-specified, which may introduce some errors.

In light of these limitations and advancements in deep learning, a possible future direction is to consider integrating classic statistical models with deep learning methods to leverage the advantages of both. By doing so, we can make more informed and precise predictions in the context of complex and evolving medical scenarios, ultimately improving patient care and decision-making.

\bibliographystyle{unsrt}  


\end{document}